\pdfoutput=1
\RequirePackage{ifpdf}
\ifpdf 
\documentclass[pdftex]{sigma}
\else
\documentclass{sigma}
\fi

\begin{document}


\renewcommand{\thefootnote}{$\star$}

\renewcommand{\PaperNumber}{093}

\FirstPageHeading

\ShortArticleName{Coherent States for Tremblay--Turbiner--Winternitz Potential}

\ArticleName{Coherent States for Tremblay--Turbiner--Winternitz\\
Potential\footnote{This
paper is a contribution to the Special Issue ``Superintegrability, Exact Solvability, and Special Functions''.
The full collection is available at
\href{http://www.emis.de/journals/SIGMA/SESSF2012.html}{http://www.emis.de/journals/SIGMA/SESSF2012.html}}}

\Author{Yusuf SUCU and Nuri UNAL}

\AuthorNameForHeading{Y.~Sucu and N.~Unal}

\Address{Department of Physics, Faculty of Science, Akdeniz University,
 07058 Antalya, Turkey}
\Email{\href{mailto:ysucu@akdeniz.edu.tr}{ysucu@akdeniz.edu.tr},
\href{mailto:nuriunal@akdeniz.edu.tr}{nuriunal@akdeniz.edu.tr}}

\ArticleDates{Received July 31, 2012, in f\/inal form November 28, 2012; Published online December 01, 2012}

\Abstract{In this study, we construct the coherent states for a
particle in the Tremblay--Turbiner--Winternitz potential by
f\/inding the conserved charge coherent states of the four harmonic
oscillators in the polar coordinates. We also derive the energy
eigenstates of the potential and show that the center of the coherent states
follow the classical orbits of the particle.}


\Classification{81R30; 81Q05; 81Q80; 81S99}

\renewcommand{\thefootnote}{\arabic{footnote}}
\setcounter{footnote}{0}

\section{Introduction}
The coherent states were derived for the one-dimensional linear
harmonic oscillator by Schr\"o\-din\-ger~\cite{1} and used in the
quantum theory of electrodynamics in 1963 and were recognized as the
Glauber states~\cite{2,3}. The coherent states are particle like localized, non-dispersive
solutions of the linear Schr\"odinger equation. In these states,
the probability density is a time dependent Gaussian wave packet and
the center of the packet follows the classical trajectory of the particle.
Recently,
the coherent states were constructed for the Kepler problem by transforming it into
four harmonic oscillators evolving in a parametric-time~\cite{4}. This technique has been applied
to derive the coherent states for a particle in the Morse potential~\cite{5},
5-dimensional Coulomb potential~\cite{6} and the
non-central generalized MIC-Kepler potential~\cite{7}. For the coherent states, the expectation
values of the position and momenta give the classical trajectories. Furthermore, the stationary
quantum eigenstates and corresponding eigenvalues may be derived from the coherent states. Therefore,
they are very important to discuss the relation between classical and quantum mechanics of a system.

Two important properties of a physical system in classical and
quantum mechanics are the exact solvability and superintegrability.
Some time ago, Tempesta, Turbiner and Winternitz conjectured that
for two dimensional systems, all maximally superintegrable systems
are exactly solvable~\cite{8}. Recently, in two dimensional plane, the following
Hamiltonian has been proposed by Tremblay, Turbiner and Winternitz
\begin{equation}
H=p_{r}^{2}+\frac{p_{\theta}^{2}}{r^{2}}+\omega^{2}r^{2}+\frac{1}{r^{2}}
\left(\frac{\alpha k^{2}}{\sin^{2}k\theta}+\frac{\beta k^{2}}{\cos^{2}k\theta}\right),
\qquad
0\leqslant\theta\leqslant\frac{\pi}{2k}
\quad\text{and}\quad
0\leqslant r<\infty,\label{1}
\end{equation}
where $\omega$ is the angular frequency of
the oscillators, $\alpha$ and $\beta$ are nonnegative constants and~$k$ is constant.
The Schr\"odinger equation has been exactly solved for this
Hamiltonian~\cite{9}. It includes all known examples of the two
dimensional superintegrable systems for $k=1,2,3$~\cite{11,10,12,13,14} or $\alpha=0$~\cite{15}.
The quantum superintegrability of the system has been also discussed
for odd~$k$ values~\cite{16}. The conjecture about the superintegrability of the
system has been supported by showing that for the classical version of
the system all bounded trajectories are closed for all integer and
rational values of the constant~$k$~\cite{17}. It has been proven that such systems in
classical mechanics are superintegrable and supported the conjecture that all
the orbits are closed for all rational values of~$k$~\cite{18}. Also,
in order to present a constructive proof for the quantum superintegrable
systems for all rational values of~$k$, the canonical operator method has been applied~\cite{19}.
In this connection, the superintegrability of a system is related into, in classical sense, the presence
of closed and periodic orbits and, in quantum mechanics sense, the presence of a non-Abelian
algebra for its integrals of motion and degenerate energy levels.

The $k=1$ case of equation~(\ref{1}) corresponds one of the four maximally
superintegrable Smorodinsky--Winternitz potentials for which coherent states have been
constructed~\cite{21,20,22}. The symmetry algebra of the system in $k=1$
case is quadratic, but, for other integer and rational values of $k$,
these systems are separable in only one coordinate frame and the symmetry al\-gebra
is not quadratic. In classical and quantum sense, the integrability properties of the system
have been investigated, in detail, for the other values of $k$, but the coherent
states hasn't been discussed so far. Since the coherent states are the macrostates as
a superposition of microstates of a system, they give both the classical and quantum
behavior of the system. Therefore, it is interesting to generalize the coherent states
and to f\/ind the expectation values of a system in equation~(\ref{1}) for the all
integer and rational values of $k$.

The aim of this study is to construct the coherent states for a
particle in a potential given by equation~(\ref{1}). In Section~\ref{section2} we f\/irst
discuss the mapping of the two dimensional system into four
dimensional system in the harmonic oscillator potential, and the
second, construct the coherent states for the four harmonic
oscillators and the third, f\/ind the coherent states for the
Hamiltonian~$H$. We also derive the energy eigenstates of the
system in polar coordinates. Section~\ref{section3}  we f\/ind the expectation values of radial and
angular coordinates of the particle in the Tremblay--Turbiner--Winternitz (TTW)
potential between the coherent states of the system. Section~\ref{section4} is our conclusions.

\section[Coherent states for Tremblay-Turbiner-Winternitz potential]{Coherent states for Tremblay--Turbiner--Winternitz\\ potential}\label{section2}

To derive the coherent states for the TTW potential, we start by writing the action~$A$
in the following way
\[
A=\int  Ldt=\int \left(p_{r}\frac{dr}{dt}+p_{\theta}\frac{d\theta}{dt}-H\right) dt,
\]
where $p_{r}$ and $p_{\theta}$ are the radial and angular momenta
of a particle in the two dimensional space with the following line element
\[
ds^{2}= (dr )^{2}+r^{2} (d\theta )^{2}.
\]
The Hamiltonian~$H$ is similar to the radial Hamiltonian of the
four harmonic oscillators. We change the angular coordinate~$\theta$ as $\Theta=k\theta$, with $0\leqslant\Theta\leqslant\frac{\pi}{2}$.
Then the Hamiltonian
becomes
\[
H=\left(p_{r}^{2}+\frac{k^{2}\mathbf{L}^{2}}{r^{2}}\right) ,
\]
where the constant of motion~$\mathbf{L}^{2}$ is given as
\[
\mathbf{L}^{2}=\left[p_{\Theta}^{2}+\left(\frac{\alpha}{\sin^{2}\Theta}+\frac{\beta}{\cos^{2}\Theta}\right)
\right].
\]

Therefore, to construct the coherent states of the system, we
f\/irst reduce the system into the $k=1$ case and take the coherent
states of the Smorodinsky--Winternitz potential. We f\/inally replace
the quantum numbers corresponding to~$\mathbf{L}^{2}$ by~$k^{2}\mathbf{L}^{2}$.

In order to represent these potentials as the kinetic energy
contribution of the angular momenta of the particle on~$S^{3}$, we
f\/irst introduce the
cartesian coordinates of the particle~$u$ and~$v$ as
\[
(u,v) =\sqrt{\omega}r(\cos\Theta,\sin\Theta) ,
\]
with the condition of $u,v\geq0$ which corresponds to represent the
system as two isotropic polar harmonic oscillators with centrifugal
barrier $\left(\frac{\alpha}{u^{2}}+\frac{\beta}{v^{2}}\right)$.
In polar coordinates $\left(\frac{\partial}{\partial u},\frac{\partial}{\partial v}\right) \Psi$
is replaced by $\left(\frac{\partial}{\partial u}+\frac{1}{2u},\frac{\partial}{\partial v}+\frac{1}{2v}\right)
\frac{\Psi}{\sqrt{uv}}$ and centrifugal barrier $\left(\frac{\alpha}{u^{2}}+\frac{\beta}{v^{2}}\right)$
become $\left(\frac{\alpha+1/4}{u^{2}}+\frac{\beta+1/4}{v^{2}}\right)$. By using
this prescription we get the following Hamiltonian in dimensionless
units
\[
\widetilde{H}=p_{u}^{2}+p_{v}^{2}+u^{2}+v^{2}+\frac{\alpha+\frac14}{u^{2}}+\frac{\beta+\frac14}{v^{2}}.
\]
Second, we introduce two dummy angles $\phi$ and~$\psi$ by the
Lagrange multipliers without changing the dynamics of the particle.
We also change $p_{r}$ and~$p_{\Theta}$ as the radial and angular
momenta of a particle in four dimensional space with the the line
element
\[
ds^{2}=(du)^{2}+(dv)^{2}+u^{2}(d\phi)^{2}+v^{2}(d\psi)^{2}.
\]
Thus the Lagrangian~$L$ becomes
\begin{equation*}
\widetilde{L}=p_{u}\frac{du}{d\omega t}+p_{v}\frac{dv}{d\omega
t}+\frac{d\phi}{d\omega t}\left(p_{\phi}-\sqrt{\alpha+\frac{1}{4}} \right) +\frac{d\psi}{d\omega
t}\left(p_{\psi}-\sqrt{\beta+\frac{1}{4}} \right) -\widetilde
{H},
\end{equation*}
where the Hamiltonian~$\widetilde{H}$ is rewritten as
\begin{equation*}
\widetilde{H}=p_{u}^{2}+p_{v}^{2}+u^{2}+v^{2}+\frac{p_{\phi}^{2}}{u^{2}}
+\frac{p_{\psi}^{2}}{v^{2}}.
\end{equation*}

The coherent states are given in conf\/iguration space~\cite{23}
$u_{i}$, $v_{i}$
with $i=1,2$ as
\[
\widetilde{\Psi}_{\lambda_{i}(t)}(u_{i})=N
 \prod\limits_{i=1}^{4}
e^{-4i\omega t}e^{-\frac{1}{2}u^{2}+\kappa_{i}(t)u_{i}}e^{-\frac{1}
{2}v^{2}+\lambda_{i}(t)v_{i}}.
\]
Here we use the natural units $\hbar=1$ and the complex
eigenvalues of the harmonic oscillators lowering operators~$\kappa_{i}(t)$ and~$\lambda_{i}(t)$ are given as
\[
\kappa_{i}(t)=\kappa_{i}(0)e^{-2i\omega t},
\qquad
\lambda_{i}(t)=\lambda_{i}(0)e^{-2i\omega t}.
\]
In polar coordinates~$(u,\phi)$ and~$(v,\psi)$
the coherent states are given as
\[
\widetilde{\Psi}_{\lambda_{i}(t)}(u_{i})
=Ne^{-4i\omega t}e^{-\frac{1}{2}u^{2}}
e^{u\left(\frac{\kappa_{1}(t)-i\kappa_{2}(t)}{2}e^{i\phi}+\frac{\kappa_{1}(t)+i\kappa_{2}(t)}{2}e^{-i\phi}\right)}
e^{v\left(\frac{\lambda_{3}(t)-i\lambda_{4}(t)}{2}e^{i\psi}
+\frac{\lambda_{1}(t)+i\lambda_{2}(t)}{2}e^{-i\psi}\right)}.
\]
If we expand the exponential expressions into the power series we get
\begin{gather}
  \widetilde{\Psi}_{\lambda_{i}(t)}(u_{i})=Ne^{-4i\omega t}e^{-\frac{1}{2}\left(u^{2}+v^{2}\right)}
\sum_{\mu_{1}=-\infty}^{+\infty} [K(0) ]^{\mu_{1}}J_{\mu_{1}}
\left(\sqrt{u^{2} (-i\kappa(t) )^{2}}\right) \exp i\mu_{1}\phi\nonumber\\
  \phantom{\widetilde{\Psi}_{\lambda_{i}(t)}(u_{i})=}
{} \times\sum_{\mu_{2}=-\infty}^{+\infty} [\Lambda(0) ]^{\mu_{2}}J_{\mu_{2}}
\left(2\sqrt{v^{2} ((-i\lambda(t)))^{2}}\right) \exp i\mu_{2}\psi, \label{K2}
\end{gather}
where $K(0)$, $\Lambda(0)$ and~$\kappa(t)$, $\lambda(t)$ are def\/ined as
\begin{alignat*}{3}
&\kappa(t)=\sqrt{[\lambda_{2}(t)+i\lambda_{1}(t)][\lambda_{2}(t)-i\lambda_{1}(t)]},\quad&&
 \lambda(t)=\sqrt{[\lambda_{4}(t)+i\lambda_{3}(t)][\lambda_{4}(t)-i\lambda_{3}(t)]},& \\
&K(0) =\sqrt{[\lambda_{2}(t)+i\lambda_{1}(t)]/ [\lambda_{2}(t)-i\lambda_{1}(t) ]},\qquad &&
 \Lambda(0) =\sqrt{[\lambda_{4}(t)+i\lambda_{3}(t)]/[\lambda_{4}(t)-i\lambda_{3}(t)]}. &
\end{alignat*}

We notice that in equation~(\ref{K2})
the eigenvalues $K(0)$ and~$\Lambda(0)$ are time independent and they correspond to the
conserved quantities (charges) $L_{12}$ and~$L_{34}$ or $p_{\phi}$
and~$p_{\psi}$. Therefore, the eigenvalues corresponding to the
polar parts of the four oscillators are time dependent.

To derive the coherent states of the physical particle~$\widetilde{\Psi}^{\text{phys}}$ we consider elimination of the
dummy coordinates~$\phi$ and~$\psi$. There are two methods of
elimination: in wave function formalism we consider physical
eigenvalues of corresponding conjugate momenta~$\widehat
{p}_{\phi}$ and~$\widehat{p}_{\psi}$ by taking care of that these
dummy coordinates are not cyclic and the conjugate momenta have
continuous eigenvalues
\[
\widehat{p}_{\phi}\widetilde{\Psi}^{\text{phys}}=\sqrt{\alpha+\frac{1}{4}}\widetilde{\Psi}^{\text{phys}},
\qquad
\widehat{p}_{\psi}\widetilde{\Psi}^{\text{phys}}=\sqrt{\beta+\frac{1}{4}}\widetilde{\Psi}^{\text{phys}}.
\]
In the path integration formalism, we integrate over all the
possible f\/inal
values of these variables. Then $\widetilde{\Psi}^{\text{phys}}$ becomes
\begin{gather}
  \widetilde{\Psi}_{\lambda_{i}(t)}^{\text{phys}}(u_{i})=Ne^{-4i\omega t}e^{-\frac{1}{2}\left(u^{2}+v^{2}\right)
}\big[K(0)e^{i\phi}\big]^{p_{\phi}}\big[\Lambda(0)e^{i\psi}\big]^{p_{\psi}} \nonumber\\
  \phantom{\widetilde{\Psi}_{\lambda_{i}(t)}^{\text{phys}}(u_{i})=}
{}\times
J_{p_{\phi}} \left(\sqrt{u^{2}(-i\kappa(t))^{2}}\right)
J_{p_{\psi}} \left(\sqrt{v^{2}(-i\lambda(t))^{2}}\right).\label{eqc}
\end{gather}
We parameterize the product of two Bessel functions as
\begin{gather*}
  J_{p_{\phi}} \left(\sqrt{u^{2}(-i\kappa(t))^{2}}\right)
  J_{p_{\psi}} \left(\sqrt{v^{2}(-i\lambda(t))^{2}}\right)\\
 \quad{}=
  J_{p_{\phi}} \left(\sqrt{u^{2}+v^{2}}\cos\Theta(-i)\sqrt{\kappa^{2}+\lambda^{2}}\cos\Phi\right)
  J_{p_{\psi}} \left(\sqrt{u^{2}+v^{2}}\sin\Theta(-i)\sqrt{\kappa^{2}+\lambda^{2}}\sin\Phi\right),
\end{gather*}
where
\[
\left(u,v\right)=\sqrt{u^{2}+v^{2}}(\sin\Theta,\cos\Theta),
\qquad
(\kappa,\lambda) =\sqrt{\kappa^{2}+\lambda^{2}}(\sin\Phi,\cos\Phi).
\]
We write the product of two Bessel functions
$J_{p_{\phi}}$ and~$J_{p_{\psi}}$ in terms of one Bessel function~\cite{24}
\begin{gather*}
  J_{p_{\phi}} \left(-i\sqrt{u^{2}+v^{2}}\sin\Theta\sqrt{\kappa^{2}+\lambda^{2}}\sin\Phi\right)
  J_{p_{\psi}} \left(-i\sqrt{u^{2}+v^{2}}\cos\Theta\sqrt{\kappa^{2}+\lambda^{2}}\cos\Phi\right) \\
 \quad{}=\sum_{l_{1}=0}^{\infty}N_{l_{1}}^{\left(p_{\phi},p_{\psi}\right)}
  \frac{J_{2l_{1}+p_{\phi}+p_{\psi}+1}\left(-i\sqrt{u^{2}+v^{2}}\sqrt {\kappa^{2}+\lambda^{2}}\right)
  }{-2i\sqrt{u^{2}+v^{2}}\sqrt{\kappa^{2}+\lambda^{2}}}
 d_{p_{\phi},p_{\psi}}^{l_{1}}(\cos2\Phi) d_{p_{\phi},p_{\psi}}^{l_{1}}(\cos2\Theta),
\end{gather*}
where the constant  $N_{l_{1}}^{(p_{\phi},p_{\psi})}$  is
\[
N_{l_{1}}^{(p_{\phi},p_{\psi})}
=\frac{i(-1)^{l_{1}}l_{1}!\Gamma(p_{\phi}+p_{\psi}+l_{1}+1)}
{\Gamma(p_{\phi}+l_{1}+1) \Gamma(p_{\phi}+l_{1}+1)},
\]
and the angular wave functions
$d_{p_{\phi},p_{\psi}}^{l_{1}}(\cos2\Theta)$
are def\/ined in terms of Jacobi polynomials \linebreak $P_{p_{\phi},p_{\psi}}^{l_{1}}(\cos2\Theta)$  as
\[
d_{p_{\phi},p_{\psi}}^{l_{1}}(\cos2\Theta)=(\sin\Theta)
^{p_{\phi}}(\cos\Theta)^{p_{\psi}}P_{p_{\phi},p_{\psi}}^{l_{1}
}(\cos2\Theta).
\]
In previous equation the generalized conserved charge coherent states are
parameterized by the time dependent eigenvalues
\[
\sqrt{\kappa^{2}+\lambda^{2}}=\sqrt{\kappa^{2}(0)+\lambda^{2}(0)}e^{-2i\omega t},
\]
and the time independent eigenvalues
\[
\sqrt{\kappa^{2}/\lambda^{2}}=\sqrt{\kappa^{2}(0)/\lambda^{2}(0)}.
\]
Here $\sqrt{\kappa^{2}/\lambda^{2}}$ corresponds to the conserved charge related to~$L^{2}$.
In equation~(\ref{eqc}) the time independent phase factors are omitted.

Then except some constant phase factors
$\widetilde{\Psi}^{\text{phys}}$
becomes
\begin{gather*}
 \widetilde{\Psi}_{\sqrt{\kappa^{2}+\lambda^{2}}}^{\text{phys}}(r,\Theta)
 =\frac{N}{4}e^{-4i\omega t}e^{-\frac{1}{2}\omega r^{2}}
 \sum_{l_{1}=0}^{\infty}N_{l_{1}}^{(p_{\phi},p_{\psi})}d_{p_{\phi},p_{\psi}}^{l_{1}}(\cos2\Phi)\\
 \phantom{\widetilde{\Psi}_{\sqrt{\kappa^{2}+\lambda^{2}}}^{\text{phys}}(r,\Theta)=}
{}\times\frac{J_{2l_{1}+p_{\phi}+p_{\psi}+1}\left(2\sqrt{\omega r^{2}}
\frac{\sqrt{(-i)^{2}\left(\kappa^{2}+\lambda^{2}\right)}}{2}\right)}
{\sqrt{\omega r^{2}}\frac{\sqrt{(-i)^{2}\left(\kappa^{2}+\lambda^{2}\right)}}{2}}
 d_{p_{\phi},p_{\psi}}^{l_{1}}(\cos 2\Theta).
\end{gather*}
Here we expand the Bessel functions in terms of Laguerre functions~\cite{25}.
The conserved charge coherent states of the four oscillators are given as
\begin{align*}
& \widetilde{\Psi}_{\sqrt{\kappa^{2}+\lambda^{2}}}^{\text{phys}}(r,\Theta)
 =\frac{N}{4}\sum_{l_{1}=0}^{\infty}N_{l_{1}}^{\left(p_{\phi},p_{\psi}\right)}
d_{p_{\phi},p_{\psi}}^{l_{1}}(\cos2\Phi)\left(\sin k\theta\right)^{p_{\phi}}
\left(\cos k\theta\right)^{p_{\psi}}P_{p_{\phi},p_{\psi}}^{l_{1}}\left(2\sin^{2}k\theta-1\right) \\
& \phantom{\widetilde{\Psi}_{\sqrt{\kappa^{2}+\lambda^{2}}}^{\text{phys}}(r,\Theta)=}
\times\sum_{n_{r}=0}^{\infty}
\frac{e^{\frac{\left(\left\vert\kappa\right\vert^{2}+\left\vert \lambda\right\vert^{2}\right)}{4}}e^{-4i\omega t}
\left(\frac{-\left(\kappa^{2}+\lambda^{2}\right)}{4}\right)^{\frac{\left(2l_{1}+p_{\phi}+p_{\psi}+1\right)}{2}+n_{r}}}
{\Gamma\left(n_{r}+2l_{1}+p_{\phi}+p_{\psi}+2\right)}\\
& \phantom{\widetilde{\Psi}_{\sqrt{\kappa^{2}+\lambda^{2}}}^{\text{phys}}(r,\Theta)=}
\times e^{-\frac{1}{2}\omega r^{2}}\left(\omega r^{2}\right)^{\frac{^{\left(2l_{1}+p_{\phi}+p_{\psi}+1\right) -1}}{2}}
L_{n_{r}}^{\left(2l_{1}+p_{\phi}+p_{\psi}+1\right)}\left(\omega r^{2}\right).
\end{align*}

In order to f\/ind the coherent states of a system described by
TTW Hamiltonian~$H$ for the terms
under the $n_{r}$ summation, we f\/irst replace $\left(2l_{1}+p_{\phi}+p_{\psi}+1\right)$
by $\left(2l_{1}+p_{\phi}+p_{\psi}+1\right) k$ and multiply the wave function
$\widetilde{\Psi}_{\sqrt{\kappa^{2}+\lambda^{2}}}^{\text{phys}}(r,\Theta)$ by $\sqrt{\omega r^{2}\sin\Theta\cos\Theta}$.

Then, we f\/ind the coherent states of the TTW system
as
\begin{gather*}
  \Psi_{\sqrt{\kappa^{2}+\lambda^{2}}}^{\text{phys}}(r,\theta)=\frac{N}
{4}\sum_{l_{1}=0}^{\infty}N_{l_{1}}^{(p_{\phi},p_{\psi})}d_{p_{\phi},p_{\psi}}^{l_{1}}(\cos2\Phi)
(\sin k\theta)^{p_{\phi}+\frac12}(\cos k\theta)^{p_{\psi}+\frac12}\nonumber\\
  \phantom{\Psi_{\sqrt{\kappa^{2}+\lambda^{2}}}^{\text{phys}}(r,\theta)=}
{} \times P_{p_{\phi},p_{\psi}}^{l_{1}}\left(2\sin^{2}k\theta - 1\right)
\! \sum_{n_{r}=0}^{\infty}\! \frac{
e^{\frac{\left(\vert\kappa\vert^{2}+\vert\lambda\vert^{2}\right)}{4}}e^{-4i\omega t}
\left(\frac{-\left(\kappa^{2}+\lambda^{2}\right)}{4}\right)
^{\frac{(2l_{1}+p_{\phi}+p_{\psi}+1)k}{4}+\frac{n_{r}}{2}}}
{\Gamma[n_{r}+(2l_{1}+p_{\phi}+p_{\psi}+1) k+1]}\nonumber\\
 \phantom{\Psi_{\sqrt{\kappa^{2}+\lambda^{2}}}^{\text{phys}}(r,\theta)=}
{}\times e^{-\frac{1}{2}\omega r^{2}}\left(\omega r^{2}\right)^{\frac{(2l_{1}+p_{\phi}+p_{\psi}+1)k}{2}}
L_{n_{r}}^{(2l_{1}+p_{\phi}+p_{\psi}+1) k}\left(\omega r^{2}\right).
\end{gather*}
Here time independent energy eigenstates are given as
\begin{gather*}
  \Psi_{n_{r},l_{1}}^{\text{phys}}(r,\theta)
 = (\sin k\theta )^{p_{\phi}+\frac12}
 (\cos k\theta )^{p_{\psi}+\frac12}P_{p_{\phi},p_{\psi}}^{l_{1}}\left(2\sin^{2}k\theta-1\right)\nonumber\\
 \phantom{\Psi_{n_{r},l_{1}}^{\text{phys}}(r,\theta)=}
 {}\times e^{-\frac{1}{2}\omega r^{2}}\left(\omega r^{2}\right)^{\frac{(2l_{1}+p_{\phi}+p_{\psi}+1)
 k}{2}}L_{n_{r}}^{(2l_{1}+p_{\phi}+p_{\psi}+1) k}\left(\omega r^{2}\right) ,
\end{gather*}
and the energy eigenvalues are
\begin{equation}
E=2(2l_{1}+p_{\phi}+p_{\psi}+1) k+2n_{r}. \label{E}
\end{equation}
The spectrum is degenerate for integer and rational values of~$k$.
So, the classical orbits are closed for these~$k$ cases.
The energy eigenstates and eigenvalues are the same with the result
given in~\cite{9}.

\section{Expectation values}\label{section3}

For the coherent states of the isotropic four harmonic oscillators the expectation values in the
conf\/iguration space $\langle u_{i}\rangle$, $\langle v_{i}\rangle$ are given as
\[
\langle u_{i}\rangle_{t}=\frac{\kappa_{i}(t)+\kappa_{i}^{\ast}
(t)}{2}=\vert \kappa_{i}(0)\vert \cos(2\omega t-\varphi_{i}) ,
\]
and
\[
\langle v_{i}\rangle_{t}=\frac{\lambda_{i}(t)+\lambda_{i}^{\ast
}(t)}{2}=\vert \lambda_{i}(0)\vert \cos(2\omega t-\chi_{i}) ,
\]
where $\varphi_{i}$ and~$\chi_{i}$ are the phases of $\kappa_{i}(0)$ and~$\lambda_{i}(0)$.
Then for the square of radial coordinate $\langle u_{i}\rangle_{t}^{2}$
the expectation value is given as
\begin{gather*}
 \langle u \rangle_{t}^{2} = \langle u_{1} \rangle_{
}^{2}+ \langle u_{2} \rangle^{2}
 =\frac{\vert \kappa_{1}(0)\vert^{2}+\vert \kappa
_{2}(0)\vert^{2}}{2}+\frac{\vert \kappa(0)\vert^{2}
}{2}\cos(4\omega t-2\varphi).
\end{gather*}
Here we have neglected the uncertainties and the constants, $\kappa(0)$
and~$2\varphi$ are given as
\[
\vert \kappa(0)\vert^{2}
=\big[\vert\kappa_{1}^{}(0)\vert^{4}+\vert \kappa_{2}(0)\vert^{4}
+2\vert\kappa_{1}(0)\kappa_{2}(0)\vert^{2}\cos2\varphi_{1}-\varphi_{2}) \big]^{\frac12},
\]
and
\[
\tan2\varphi
=\frac{\vert \kappa_{1}(0)\vert^{2}\sin2\varphi_{1}+\vert \kappa_{1}(0)\vert^{2}\sin2\varphi_{2}}
{\vert \kappa_{1}(0)\vert^{2}\cos2\varphi_{1}+\vert\kappa_{1}(0)\vert^{2}\cos2\varphi_{2}}.
\]
In order to derive the expectation values between the conserved charge coherent
states, we evaluate the angular momentum $L_{12}$ as
\begin{gather*}
L_{12} =-\frac{\left(\kappa_{1}(t)\kappa_{2}^{\ast}(t)-\kappa_{2}(t)\kappa_{1}^{\ast}(t)\right)}{2i}
 =-\left\vert \kappa_{1}(0)\kappa_{2}(0)\right\vert \sin\left(\varphi_{1}-\varphi_{2}\right).
\end{gather*}
We f\/ix the value of $L_{12}$ as $\sqrt{\alpha+\frac{1}{4}}$. Then
\[
\frac{\cos2(\varphi_{1}-\varphi_{2})}{2}
=1-\frac{2\left(\alpha+\frac{1}{4}\right)}{\vert \kappa_{1}(0)\kappa_{2}(0)\vert^{2}},
\]
and
\[
\vert \kappa(0)\vert^{2}
=\left[\big(\vert\kappa_{1}(0)\vert^{2}+\vert \kappa_{2}(0)\vert^{2}\big)^{2}
-4\left(\alpha+\frac{1}{4}\right) \right]^{\frac12}.
\]
As a result of this, for $k=1$, $\langle u\rangle_{t}^{2}$ is
given as
\begin{gather}
\langle u \rangle_{t}^{2}
  = \langle u_{1} \rangle^{2}+ \langle u_{2} \rangle^{2}\nonumber\\
\hphantom{\langle u \rangle_{t}^{2}}{}
 =\frac{\vert \kappa_{1}(0)\vert^{2}+\vert\kappa_{2}(0)\vert^{2}}{2}
+\left[\frac{\big(\vert \kappa_{1}(0)\vert^{2}
+\vert \kappa_{2}(0)\vert^{2}\big)^{2}}{4}-\left(\alpha+\frac{1}{4}\right) \right]^{\frac12}
\cos(4\omega t-2\varphi).\label{Equ}
\end{gather}
In the similar way $\langle v\rangle_{t}^{2}$ is given as
\begin{gather*}
 \langle v \rangle_{t}^{2}
  = \langle v_{1} \rangle^{2}+ \langle v_{2} \rangle^{2}\nonumber\\
\hphantom{\langle v \rangle_{t}^{2}}{}
 =\frac{\vert \lambda_{1}(0)\vert^{2}+\vert\lambda_{2}(0)\vert^{2}}{2}
+\left[\frac{\big(\vert \lambda_{1}(0)\vert^{2}
+\vert \lambda_{2}(0)\vert^{2}\big)^{2}}{4}-\left(\beta+\frac{1}{4}\right) \right]^{\frac12}
\cos (4\omega t-2\chi ).
\end{gather*}
These are the expectation values of $\langle u\rangle_{t}^{2}$
and~$\langle v\rangle_{t}^{2}$ between the conserved charge coherent
states for the radial oscillators given by the Hamiltonian~$\widetilde{H}$.
In two-dimensional oscillator with polar coordinates $(u,\phi)$ the expectation value of unit vector~$\big\langle\widehat\phi \big\rangle$ also oscillates in time.
For the reduction of dimension we f\/ix the oscillations in
$ \langle\widehat\phi  \rangle$ and take it as a constant unit vector.

To f\/ind the expectation value of radial coordinates we arrange the following
product
\begin{gather*}
 \left[\kappa_{1}^{2}(t)+\kappa_{2}^{2}(t)+\lambda_{1}^{2}(t)+\lambda_{2}^{2}(t)\right]
 \left[\kappa_{1}^{\ast2}(t)+\kappa_{2}^{\ast2}(t)+\lambda_{1}^{\ast2}(t)+\lambda_{2}^{\ast2}(t)\right] \\
 \qquad
=\left[ \vert\kappa_{1}(0) \vert^{2}+ \vert\kappa_{2}(0) \vert^{2}
+ \vert \lambda_{1}(0) \vert^{2}+ \vert \lambda_{2}(0) \vert^{2}\right]^{2}\\
 \quad\qquad{}
+\left[\kappa_{1}(0)\kappa_{2}^{\ast}(0)-\kappa_{2}(0)\kappa_{1}^{\ast}(0)\right]^{2}
+\left[\lambda_{1}(0)\lambda_{2}^{\ast}(0)-\lambda_{2}(0)\lambda_{1}^{\ast}(0)\right]^{2}\\
 \quad\qquad{}
+\left[\kappa_{1}(0)\lambda_{1}^{\ast}(0)-\lambda_{1}(0)\kappa_{1}^{\ast}(0)\right]^{2}
+\left[\kappa_{1}(0)\lambda_{2}^{\ast}(0)-\lambda_{2}(0)\kappa_{1}^{\ast}(0)\right]^{2}\\
 \quad\qquad{}
+\left[\kappa_{2}(0)\lambda_{1}^{\ast}(0)-\lambda_{1}(0)\kappa_{2}^{\ast}(0)\right]^{2}
+\left[\kappa_{2}(0)\lambda_{2}^{\ast}(0)-\lambda_{2}(0)\kappa_{2}^{\ast}(0)\right]^{2}\\
 \qquad{}
=4\left(\frac{E}{\omega}\right)^{2}-4\mathbf{L}^{2}.
\end{gather*}
Then, for $k=1$ case, $\langle r\rangle_{t}^{2}$ becomes
\begin{equation}
\omega\langle r\rangle_{t}^{2}=\left(\frac{E}{2\omega}\right)
+\left[\left(\frac{E}{2\omega}\right)^{2}-\mathbf{L}^{2}\right]^{\frac12}\sin2(2\omega t-\delta).\label{Eqr}
\end{equation}
Here the constants of motion are def\/ined as
\[
\frac{E}{\omega}=\frac{\vert \kappa_{1}(0)\vert^{2}
+\vert \kappa_{2}(0)\vert^{2}+\vert \lambda_{1}^{{}
}(0)\vert^{2}+\vert \lambda_{2}(0)\vert^{2}}{2},
\]
and
\[
\cot2\delta=\frac{\vert \kappa(0)\vert^{2}\sin2\varphi+\vert
\lambda(0)\vert^{2}\sin2\chi}{\vert \kappa(0)\vert^{2}
\cos2\varphi+\vert \lambda(0)\vert^{2}\cos2\chi}.
\]
In order to derive the expectation values for TTW
Hamiltonian we replace $\mathbf{L}^{2}$, $\alpha+\frac12$, and $\beta+\frac12$ by
$k^{2}\mathbf{L}^{2}$, $k\left(\alpha+\frac12\right)$ and $k\left(\beta+\frac12\right)$, respectively.
Then, $\omega\langle r\rangle_{t}^{2}$ becomes
\[
\langle r\rangle_{t}^{2}
=\frac{E}{2\omega^{2}}+\left[\left(\frac{E}{2\omega^{2}}\right)^{2}-\frac{A}{\omega^{2}}\right]^{\frac12}
\sin4\omega(t-t_{0}),
\]
where $A$ is $k^{2}\mathbf{L}^{2}$. The $\langle r\rangle_{t}$
shows the center of the coherent states following the classical trajectories of
the particle in this TTW potential. It should be emphasized that the result on
the radial coordinate of expectation value derived from the coherent states agrees
with the result of~\cite{17} in which bounded trajectories are investigated for
the integer and rational values of~$k$.

The expectation value of $\sin\Theta$ may be derived from
$\langle u\rangle_{t}^{2}$ and $\omega\langle r\rangle_{t}^{2}$ in equations~(\ref{Equ})
\mbox{and~\eqref{Eqr} as}
\[
\big\langle \sin\Theta\big\rangle^{2}
=\left.\frac{\langle u\rangle_{t}^{2}}{\omega\langle r\rangle_{t}^{2}}
\right|_{L_{12}^{2}=k\left(\alpha+\frac12\right)}.
\]

\section{Conclusion}\label{section4}

In this study, we have constructed the coherent states for the
Tremblay--Turbiner--Winternitz potential. One of the constants of
motion is~$k^{2}\mathbf{L}^{2}$. Therefore, for the case $k\neq1$,
we have integrated the equations of motion in polar coordinates by
using the solutions for $k=1$ case, since the separability in this
coordinates do not depend on the value of~$k$. For $k=1$ case, the
Schr\"odinger equation is also separable in cartesian coordinates~$(u,v)$.

For the Tremblay--Turbiner--Winternitz potential, the coherent
states correspond to the conserved charge coherent states of the
four oscillators with the eigenvalues of
$k\sqrt{\alpha+\frac{1}{4}}$, $k\sqrt{\beta+\frac {1}{4}}$ and
$k^{2}\left[2l_{1}(2l_{1}+p_{\phi}+p_{\psi}+1)\right]$ of the operators $kL_{12}$, $kL_{34}$,
and $k^{2}\left[p_{\Theta}^{2}+\left(\frac{\alpha}{\sin^{2}\Theta}+\frac{\beta}{\cos^{2}\Theta}\right)
\right]$, respectively. The energy eigenvalues are given as by
equation~(\ref{E}), and as it was shown previously in~\cite{9}, the
spectrum becomes degenerate for the integer and rational values of
the parameter~$k$. This result corresponds to the periodicity of
the Hamiltonian: if the Hamiltonian has angular periodicity of
$\frac{\pi m}{n}$ for any of two integers~$m$ and~$n$, then~$k$
must be an integer or rational number. Furthermore, we have evaluated
the expectation value of the radial coordinate~$\langle r\rangle_{t}$,
and shown that the center of the coherent states follow the classical trajectories
of the particle in the Tremblay--Turbiner--Winternitz potential.

\subsection*{Acknowledgements}

This work was supported by Akdeniz University, Scientif\/ic Research
Projects Unit.

\pdfbookmark[1]{References}{ref}
\LastPageEnding

\end{document}